%% file: main.tex
\documentclass{article}
\usepackage{smc}
\usepackage{times}
\usepackage{ifpdf}
\usepackage[english]{babel}
\usepackage{cite}
\usepackage{amsfonts}
\usepackage{amsmath}
\usepackage{multirow}

\usepackage{dirtytalk} 

\def\papertitle{A Machine Learning Approach for MIDI to Guitar Tablature Conversion}
\def\firstauthor{First author}
\def\secondauthor{Second author}
\def\thirdauthor{Third author}

\newif\ifpdf
\ifx\pdfoutput\relax
\else
   \ifcase\pdfoutput
      \pdffalse
   \else
      \pdftrue
   \fi                 
\fi                    

\ifpdf 
  \usepackage[pdftex,
    pdftitle={\papertitle},
    pdfauthor={\firstauthor, \secondauthor, \thirdauthor},
    bookmarksnumbered, 
    pdfstartview=XYZ 
   ]{hyperref}

  \usepackage[pdftex]{graphicx}
  \graphicspath{{./figures/}}
  \DeclareGraphicsExtensions{.pdf,.jpeg,.png}

  \usepackage[figure,table]{hypcap}

\else 
  \usepackage[dvips,
    bookmarksnumbered, 
    pdfstartview=XYZ 
  ]{hyperref}  

  \usepackage[dvips]{epsfig,graphicx}
  \graphicspath{{./figures/}}
  \DeclareGraphicsExtensions{.eps}

  \usepackage[figure,table]{hypcap}
\fi

\hypersetup{
    colorlinks,%
    citecolor=black,%
    filecolor=black,%
    linkcolor=black,%
    urlcolor=black
}

\title{\papertitle}

\oneauthor
{%
  \parbox{\linewidth}{%
    \centering
    Maximos Kaliakatsos-Papakostas$^{1}$, Gregoris Bastas$^{1,2}$,
    Dimos Makris$^{3}$, Dorien Herremans$^{3}$,\\
    Vassilis Katsouros$^{1}$, Petros Maragos$^{2}$
  }%
}
{%
  $^{1}$ Institute for Language and Speech Processing, Athena R.C., Athens, Greece\\
  $^{2}$ School of Electrical and Computer Engineering, NTUA, Athens, Greece\\
  $^{3}$ Department of Computer Science and Design Pillar, SUTD, Singapore\\
  \texttt{\{maximos,g.bastas,vsk\}@athenarc.gr}, \texttt{maragos@cs.ntua.gr}\\
  \texttt{\{dimosthenis\_makris,dorien\_herremans\}@sutd.edu.sg}
}

\begin{document}
\capstartfalse
\maketitle
\capstarttrue
\begin{abstract}
Guitar tablature transcription consists in deducing the string and the fret number on which each note should be played to reproduce the actual musical part. This assignment should lead to playable string-fret combinations throughout the entire track and, in general, preserve parsimonious motion between successive combinations. Throughout the history of guitar playing, specific chord fingerings have been developed across different musical styles that facilitate common idiomatic voicing combinations and motion between them. This paper presents a method for assigning guitar tablature notation to a given MIDI-based musical part (possibly consisting of multiple polyphonic tracks), i.e.\ no information about guitar-idiomatic expressional characteristics is involved (e.g.\ bending etc.) The current strategy is based on machine learning and requires a basic assumption about how much fingers can stretch on a fretboard; only standard 6-string guitar tuning is examined. The proposed method also examines the transcription of music pieces that was not meant to be played or could not possibly be played by a guitar (e.g.\ potentially a symphonic orchestra part), employing a rudimentary method for augmenting musical information and training/testing the system with artificial data. The results present interesting aspects about what the system can achieve when trained on the initial and augmented dataset, showing that the training with augmented data improves the performance even in simple, e.g.\ monophonic, cases. Results also indicate weaknesses and lead to useful conclusions about possible improvements.
\end{abstract}

\input{0_intro}

\input{1_method}

\input{2_data}

\input{3_results}

\input{4_conclusions}

\begin{acknowledgments}
This research has been co-financed by the European Regional Development Fund of the European Union and Greek national funds through the Operational Program Competitiveness, Entrepreneurship and Innovation, under the call RESEARCH - CREATE - INNOVATE (NLP-Theatre, project code: T1EDK-00508). This work is supported by Singapore Ministry of Education Grant no. MOE2018-T2-2-161.
\end{acknowledgments} 

\bibliography{smc2022bib}

\end{document}

%% file: 0_intro.tex
\section{Introduction}\label{sec:intro}

Tablature constitutes a very old form of music notational language which provides guidance towards an active and intimate interconnection with the instrument, in contrast to music scores which aim at capturing more abstract psychoacoustic characteristics of sound, such as pitch or intervals \cite{kojs2011notating}. 
A common type of tablature these days is guitar tablature (hereinafter simply referred to as \say{tablature}) and represents sequences of string-fret combinations that correspond to specific notes. It contains information about the fretboard positions on which the notes are articulated, which is not contained in typical music scores. However, tablature notation has limited expressive power when it comes to rhythmic structure and dynamics. It is this combination of features, though, which renders tablatures valuable in many occasions, since, assuming some auditory familiarity with the represented  piece, they are more easily readable by non-experts. 
This can be important for general teaching purposes, especially among beginner and self-taught guitar players.
Additionally, style-specific requirements have been established that would guide a transcriber of a piece to specific fingering choices, e.g., preference for lower-pitches strings when playing power-chords in metal music. 
Thus, a system that can generate tablatures, given the notes that make up a musical work, can be of great significance for the guitarist community. The difficulty of this task mainly arises due to the guitar's design, since same pitch notes can be played on more than one position on the fretboard. 
These facts make machine learning approaches relevant: how can a system learn from context what the best tablature is for a given set of pitches?



Related research includes generating tablature notation directly from audio, midi, or sheet music.
Some research has focused on transcribing tablature directly from audio by developing accurate string classification models. Barbancho et al. \cite{barbancho2012inharmonicity} proposed a method to simultaneously detect the pitch and string on which a note should best be played by relying on inharmonicity analysis that enabled accurate partial tracking. Other research teams have used estimations of the inharmonicity coefficient in order to train bayesian statistical models
 \cite{michelson2018automatic, hjerrild2019estimation}. The inharmonicity coefficient has also been used as one of the input features to classification models, such as Support Vector Machines (SVMs) \cite{abesser2012automatic, kehling2014automatic}.
%
The above strategies focus mainly on monophonic performances with the exception of \cite{barbancho2012inharmonicity} who incorporate chord recognition within certain limits.


Some research on tablature transcription has focused on deducing plausible sequences of string-fret combinations, relying mostly on playability constraints. They dispose of string-specific audio features, which are difficult to  acquire in polyphonic performances, and practically reduce the problem to two phases: (multi-)pitch detection followed by pitch to tablature conversion. 
Barbancho et. al \cite{barbancho2011automatic} proposed a system for chord and fingering recognition using Hidden Markov Models (HMMs) that encode the probabilities to move from one configuration to the other. Burlet and Fujinaga \cite{burlet2013robotaba} encoded string-fret transitions using weighted directed acyclic graphs (WDAGs) and the $A^*$ algorithm was employed to determine the optimal sequences. In \cite{yazawa2013audio} and \cite{yazawa2014automatic}, where the audio was produced by MIDI, a graph-based representation was employed using WDAGs, and a dynamic programming (DP) algorithm was applied to find the longest paths. 
	
Several other research teams have worked on fingering analysis, playable guitar tablature generation, and fingering recognition given a series of note events in symbolic notation (music scores or MIDI) either by employing graph representations, Dynamic Programming (DP) strategies \cite{sayegh1989fingering,radisavljevic2004path,radicioni2005guitar,miura2004constructing,hori2013input}, or other optimization algorithms  \cite{tuohy2005genetic,tuohy2006ga,tuohy2006evolved,tuohy2006generating,tuohy2006guitar,ramos2015comparative,ramos2016evaluation}. For example,
Tuohy and Potter \cite{tuohy2005genetic} proposed a simple genetic algorithm (GA) which incorporates limitations related to guitar playability in a fitness function. The same team has managed to combine Artificial Neural Networks (ANNs) with a GA \cite{tuohy2006evolved} for the problem of musical score to tablature conversion. In \cite{tuohy2006generating} they also ascribe fingerings to the predicted string-fret positions. 	


ANNs have also been employed to tackle the problem of tablature estimation directly from audio signals. Multi-layer Perceptrons \cite{gagnon2003neural} and Convolutional Neural Networks (CNNs) \cite{humphrey2014music} have been used for chord recognition and the detection of hand positions corresponding to chords.
More recently, Wiggins and Kim \cite{wiggins2019guitar} applied CNNs for string-fret recognition in solo performances that which also included elements of polyphony. Deep learning methods were also recently applied  as a method for guitar tablature composition (generation) \cite{chenautomatic}, more specifically, using a Transformer.


%% file: 1_method.tex
\section{Method}\label{sec:method}



The proposed method\footnote{\url{https://github.com/maximoskp/midi2tab\_Deconvolutions}} assumes normal-tuning guitars with six strings and 24 frets -- 25 values per string are considered for representing the free string as well. The transcription is considered to involve only information about what midi pitches need to be transcribed, disregarding information about inter onset intervals between successive sets (or frames) of pitches to be transcribed; also midi velocity and duration are not considered. Therefore, the transcription task is described as a ``binary process'', where a binary representation of input midi pitches is transcribed into a binary representation of a tablature. The overall method is divided in two parts for reaching a single tablature decision: a) a deep neural network is employed for generating a ``probabilistic'' tablature that indicates which string-fret combinations are more probable for a given input and b) a search algorithm that uses a simple conceptualisation of guitar ``playability'' for the best string-fret positions given the previous finger positions and taking into account the highest-probability tablature frame.

The aim of the overall method is to assign a playable tablature notation to a given set of notes (midi), whether or not the complete set of these notes is playable on a guitar fretboard, e.g.\, because there are more than six simultaneous notes or because the notes stretch over a wide octave range that cannot be covered by a guitar. During the application of the algorithm in the second part of the method, which involves examining all fretboard combinations that implement the input midi pitches, it is possible to iteratively discard midi pitches from being considered for transcription. This can either happen because the midi input frame might involve more than six pitches (which are impossible to be played simultaneously in a six-string guitar) or because of improper pitch layout (e.g.\, two pitches that correspond to the 1st and 2nd string fret of the lower E string). It is also assumed that guitar fingering on the tablature does not only depend on the current (midi) pitches to be played, but also from previous finger positions on the fretboard, or previous fretboard frames. Motion of the fingers on the fretboard between successive tablature frames, accounts for minimizing the travel distances across the fretboard and for preserving fingering shapes between successive frames as much as possible.

\subsection{Probabilistic Fretboard Deep Neural Network}

\begin{figure}[t]
\centering
\includegraphics[width=0.45\textwidth]{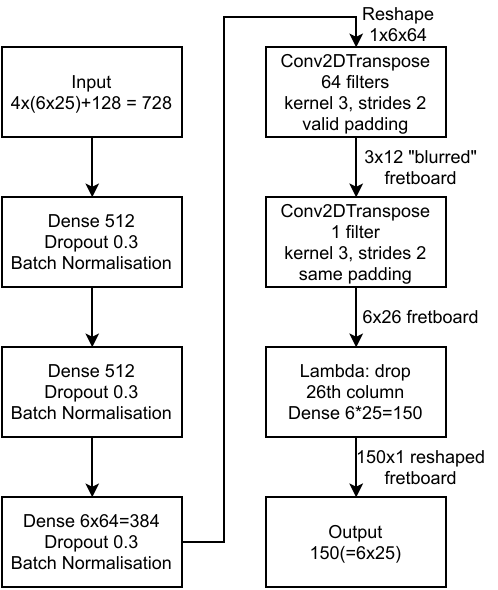}
\caption{Examined architecture.\label{fig:cnn_architecture}}
\end{figure}

The deep neural network that was developed for solving the problem of assigning tablature to midi input is a minimal approach towards modeling fingering shapes and time dependencies between successive tablature frames. The architecture is shown in Figure~\ref{fig:cnn_architecture}. Fingering patterns are expected to be learned from data. A Convolutional Neural Network (CNN) architecture is chosen for doing so, since convolutional filters are designed to capture spatial patterns, e.g.\, on a 2D fretboard. Specifically, the output of the network employs two layers of transposed convolutions (or deconvolutions) that result in the formation of a $6\times 26$ fretboard with 6 string and 25 frets plus the free string (there is a lambda layer dedicated to simply dropping the 25th fret information). The first layer of deconvolutions (top right box in Figure~\ref{fig:cnn_architecture}) generates a $3\times 12$ structure that can be thought of as ``blurry'', or highly pixelated, low resolution version of a fretboard image. This layer learns to describe filter combinations in the next deconvolution layer, which actually forms the final version of the aforementioned $6\times 26$ fretboard.

Input to these two deconvolutional layers is a ``latent'' space with 384 dimensions, which is formed by successive transformations of three feedforward layers (with scaled exponential linear activation) of the overall input to the network. The input of the network includes not only the midi pitches to be transcribed to tablature, but also the 4 previous frames of the tablature, as a crude approach to involving information about successive frame dependencies. The input is a concatenated array + that incorporates this information, of total size 728: 128 midi pitches to be transcribed plus 600 (4 times 125) for the four previous tablature frames. All input is a binary vector, with ones representing the index of active midi pitches to be transcribed and active string-fret combinations in each input tablature frame.

During the training phase, the network learns to approximate the target tablature (in binary representation) with output activity of the deconvolutional layers. Even though no encoder-decoder parts are explicitly constructed, it might be helpful to think of the vertical alignment of boxes on the left part of Figure~\ref{fig:cnn_architecture}) as the encoder, which generates a latent representation of the input with 384 elements, and the right part as the decoder, which generates a ``\textit{probabilistic tablature}'' based on the latent space generated by the input. The term ``probabilistic tablature'' is employed for denoting the fact that the network is able to generate an approximation of the tablature that is not binary; i.e.\ it incorporates real values that indicate (without forming a probability density function, that sums to one) the probability of a string-fret pair being active.

\subsection{Probabilistic Fretboard to Actual Fingering}\label{subsec:prob2actual}

This section describes the process of assigning actual fingerings based on the ``probabilistic fretboard'' produced by the network. The algorithm for keeping all possible and playable fretboards from a set of midi pitches $\vec{m}$ is summarised as follows:
\begin{enumerate}
    \item Bring all pitches of $\vec{m}$ within the fretboard range, by increasing or decreasing their pitch values by an octave.
    \item Decide on the number of pitches from $\vec{m}$ to be preserved, which is either the number of pitches $\vec{m}$ or 6, if this number exceeds 6 (since only 6-string guitars are considered). This number is symbolized as $N$.
    \item\label{item:combinations} Keep all combinations of pitches of $\vec{m}$ that include $N$ notes.
    \item\label{item:playable} For each combination, generate all possible binary fretboards and keep the ones that are ``\textit{playable}''. A binary fretboard is considered to be \textit{playable} if: a) there is at most one pitch per string and b) all non-open string pitches fall within a window of 6 frets (considering that finger stretching of over 6 frets is not acceptable).
    \item If there is no binary fretboard that satisfies the above mentioned criteria, reduce $N$ by one and go back to step~\ref{item:combinations}.
\end{enumerate}

\begin{figure}[h]
\centering
\includegraphics[width=0.49\textwidth]{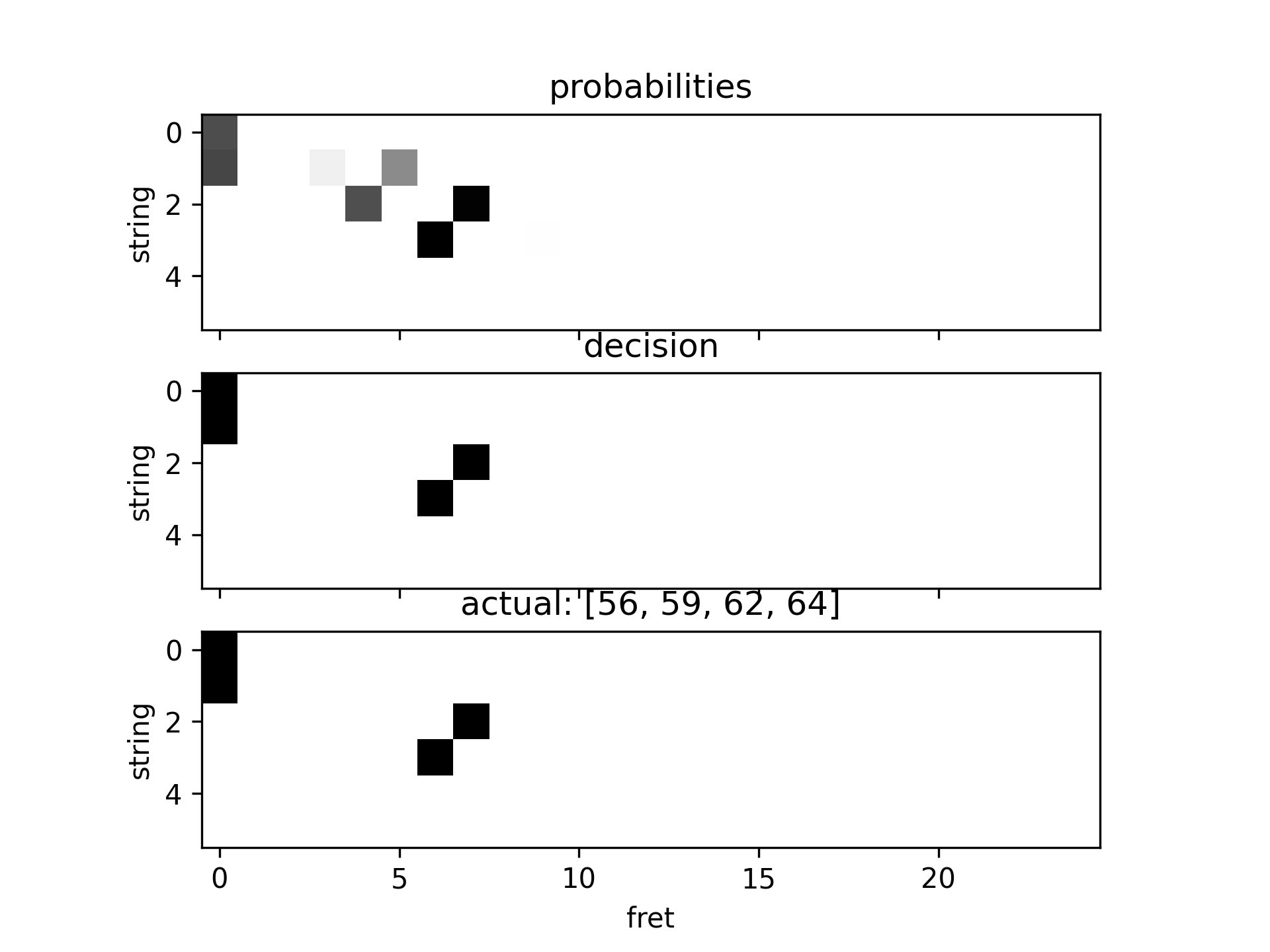}
\caption{Examined architecture.\label{fig:probs2bin_example}}
\end{figure}

After all possible binary fretboards are obtained, denoted as $b_k\in \{0,1\}^{6\times 25}$, $k\in \{1, 2, \dots , K\}$, where $K$ is the number of retrieved possible binary fretboards, the fretboard that corresponds to the higher sum of values in the probabilistic tablature provided by the network ($p\in \mathbb{R}^{6\times 25}$) is retained. This sum is obtained through the inner product of the probabilistic tablature and each examined binary tablature:
\begin{equation}\label{eq:analytic}
    b = \operatorname*{argmax}_k (p\cdot b_k)\text{,}
\end{equation}
where $\cdot$ denotes the inner product of $b_k$ and $p$. Figure~\ref{fig:probs2bin_example} depicts the decision-making process, where the top graph shows the probabilistic tablature generated by the network, the bottom shows the ground-truth tablature and the middle part shows the binary tablature, among all the possible ones for the pitches to be transcribed, that achieves maximum value of the inner product in eq.~\ref{eq:cosine}. The selected binary tablature (middle) is the same as the ground-truth tablature (bottom) in this case.

%% file: 2_data.tex
\section{Dataset and data manipulation}\label{sec:data}

The DadaGP dataset~\cite{sarmento2021dadagp} was used as dataset for the current paper. This dataset is comprised of 26,181 pieces in GuitarPro (GP) format, i.e.\ in tablature notation for guitar and string instruments, and those pieces are divided into 739 genres. Even though an encoded form of the dataset in tokenized form is also available, the study at hand employs the original GP files for the necessary conversions presented in this section. Data in the system was represented in a binary format (described in the next paragraph) that generated multiple instances per piece, making the entirety of the dataset practically unusable because of its large size. Therefore, a 5\% part of the entire dataset was employed, in which a random 5\% of pieces from each folder in the DadaGP was included. Since the folder structure in DadaGP divides pieces per album, it is expected that this data selection method creates a set with a distribution among all available styles that is similar to the distribution in the original set. Additionally, the approach presented herein discusses only transcription for 6-string guitars in standard tuning, excluding all piece that employ different tunings and number of strings. This study also excludes guitar-idiomatic and expressional information, e.g.\ string bending, hammer-on, pull-off etc. Even though this information is crucial for guitar transcription, it is left for future work.

From the pieces that are retained, only the guitar tracks are preserved. From each guitar track, discrete successive events are extracted, where each event includes pitch-related fretboard information for any change that occurs on the fretboard. Each event represents an instance of a fretboard layout at a given time, where potentially multiple notes are active, each note on a separate string (i.e., at most 6 notes can be played simultaneously). A fretboard layout, in turn, is represented as a 6-by-25 binary matrix, where rows represent strings and columns 24 (plus a free string) frets. In each binary matrix, ones indicate which pitch is played on which fret (first columns of each string indicate a free string); each binary matrix is hereby referred to as a `` tablature frame'' and each piece includes so many frames as the number of tablature changes it includes. Each tablature frame is then transcribed to the respective binary midi frame; primary aim of the presented method is to develop a method for translating midi frames to tablature frames.

\begin{figure*}[ht]
\centering
\includegraphics[width=0.95\textwidth]{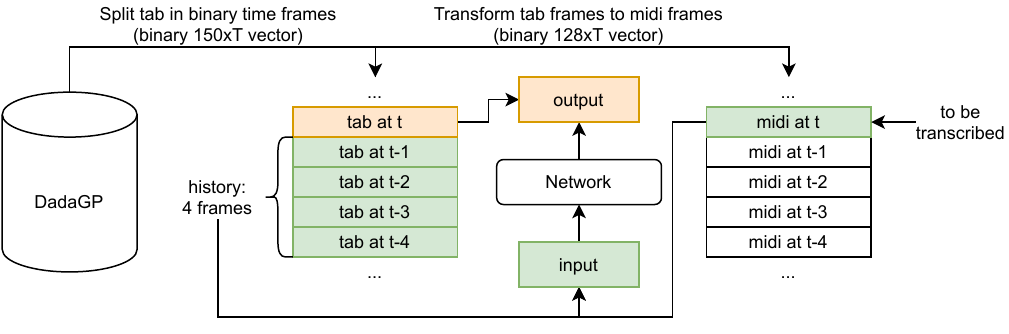}
\caption{Data preparation overview.\label{fig:data_preparation}}
\end{figure*}

Tablature frames are reshaped to row-shaped arrays with 150 ($6\times 25$) elements and each piece is represented as a stack of successive frames on top of each other; simlarly with midi frames. These two stacks of frames are shown in Figure~\ref{fig:data_preparation}. Each piece is also padded with a number of all-zero frames, depending on the assumed tablature ``history'' considered in the model, i.e., how many previous frames should be considered for identifying fingering patterns that are more probable at any given instance. Tablature history is assumed an important factor when deciding what the next fingering layout should be. Therefore, tablature history is included in the design of the examined model. To this end, and according to Figure~\ref{fig:data_preparation}, considering a history of four frames, the input to the system is the tablature included in the past four frames (600 binary numbers) along with the current midi to be transcribed in the current frame (128 binary numbers, totalling an input of 728 numbers). The system learns to generate output that predicts the current tablature frame. Summarising, the model learns to transform an input of 728 binary digits to 150 binary digits. The 5\% of the DadaGP dataset that is retained totals a number of 955971 training, 239402 validation and 299423 test time frames.

Another aim is to study whether the presented method is flexible enough to transcribe midi input that is not playable by a single guitar. Since there is no dataset incorporating such information, a naive music-driven data alteration method is followed, which ``augments'' midi information. This augmentation (in terms of musical information) method assumes that any tablature frame might have been a result of transcribing richer (in terms of number of pitches) midi content. To this end, guitar frame transformation from tablature to midi involves the insertion of additional pitches. Specifically, for each pitch in the tablature, there is a 50\% chance that a new midi pitch will be inserted an octave up, down, a fifth, a fourth, a major/minor third or a major/minor sixth (selecting uniformly between these notes). The midi input will therefore include more pitches than the ones that appear in the target tablature, forcing the network to learn to ignore midi pitches that do not fit into the context of guitar transcription.

The intention is to make the system selective towards tablature interpretations of midi input that are more convenient for guitar playability, allowing the possibility to exclude pitches that do not offer much in terms of musical information and, at the same time, would possibly result in fingerings that are less frequent in the dataset. This data augmentation method has severe weaknesses, especially in cases where the tablature frame involves few notes. For instance, if the tablature frame includes one pitch, then augmenting the respective midi frame with, e.g., a major third up would create a pair of notes that would be easily playable with a guitar and, in fact, there are still multiple instances of this pair of notes in the original, not augmented dataset. This potentially leads to multiple conflicting tablature outputs for the same midi input, where in one case, the non-augmented midi input would correspond to the exact fingering in the target tablature, while in another case, the augmented midi output would (possibly) be the same as in the first case, but would correspond to a different tablature. This method could be further refined, e.g., to allow augmentation only in cases where a number of simultaneous pitches above a predefined threshold is involved (e.g., augmentation could occur to tablature frames that include more than four notes). The simplest approach, however, that possibly adds pitches in any frame, presents results that are interesting, so this augmentation method is used in the reported results in this paper.

%% file: 3_results.tex
\section{Results}\label{sec:results}

The method is examined in terms of learning adaptivity for the neural network and in terms of actual transcription characteristics for the entire system. The learning errors of the neural network provide an indication about how well the ``probabilistic tablature'' output generated by the network approaches the target tablatures as the epochs increase in terms of the mean square error, but these errors do not provide an intuitive assessment of how clearly this output can be translated into the binary target tablatures at each epoch. Ideally, a complete binary output of the system should be employed to evaluate the accuracy of the system, however, it is prohibitively slow to apply the analytical step for deciding the best binary tablature for any given probabilistic tablature (input data point), in each epoch, in the training and the evaluation steps.

To this end, a fast to compute and intuitive measurement of how well the probabilistic tablature output is aligned with the target binary output is the cosine similarity, given by
\begin{equation}\label{eq:cosine}
    \cos({\theta }) = \frac{\vec{p}\cdot \vec{b}}{\|\vec{p} \| \| \vec{b} \|} \text{,}
\end{equation}
where $\vec{p}$ is the probabilistic tablature generated by the network and $\vec{b}$ is the target binary tablature. This measure computes the dot product over the product of the norms of each vector, showing, intuitively, how well the predictions are aligned (or correlated) with the (binary) target, regardless of the magnitude in these predictions. In the analytical step of the tablature generation process, the $\vec{p}$ produced by the network is compared with all possible binary tablatures $\vec{b_k}$ through the dot product (eq. \ref{eq:analytic}). The difference between eq. \ref{eq:analytic} and eq. \ref{eq:cosine} is that in the latter, the inner product is normalized with the norm of both involved vectors, while in the former this is not necessary, since the maximum value is assessed and all binary tablature vectors have the same norm (all include the same number of ones).

\begin{figure*}[ht]
   \centering
\begin{tabular}{cc}
\includegraphics[width=0.45\textwidth]{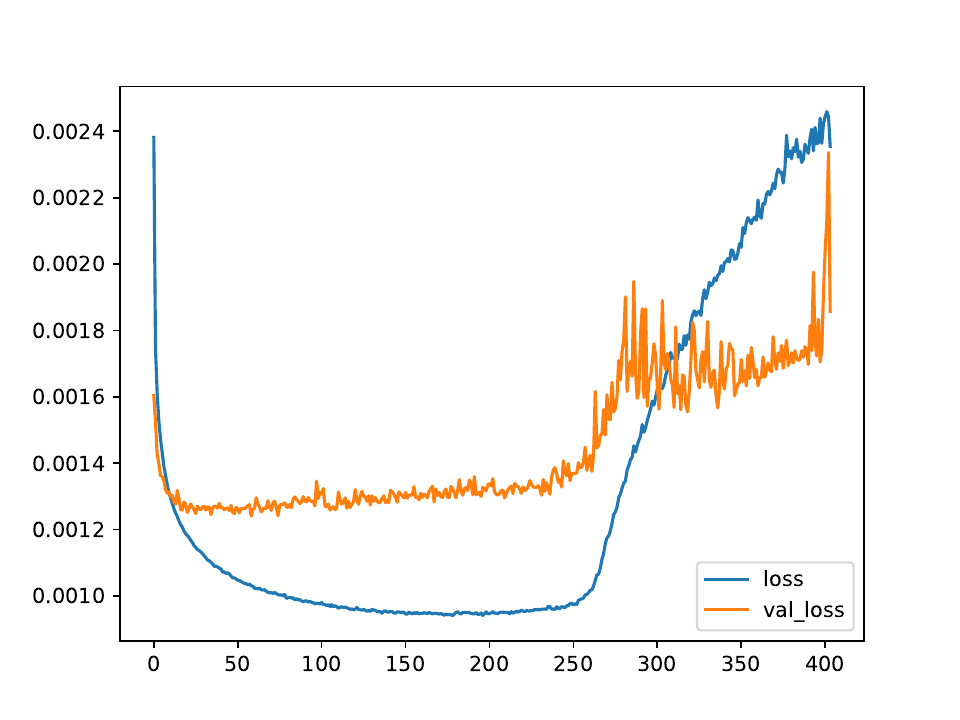}&
\includegraphics[width=0.45\textwidth]{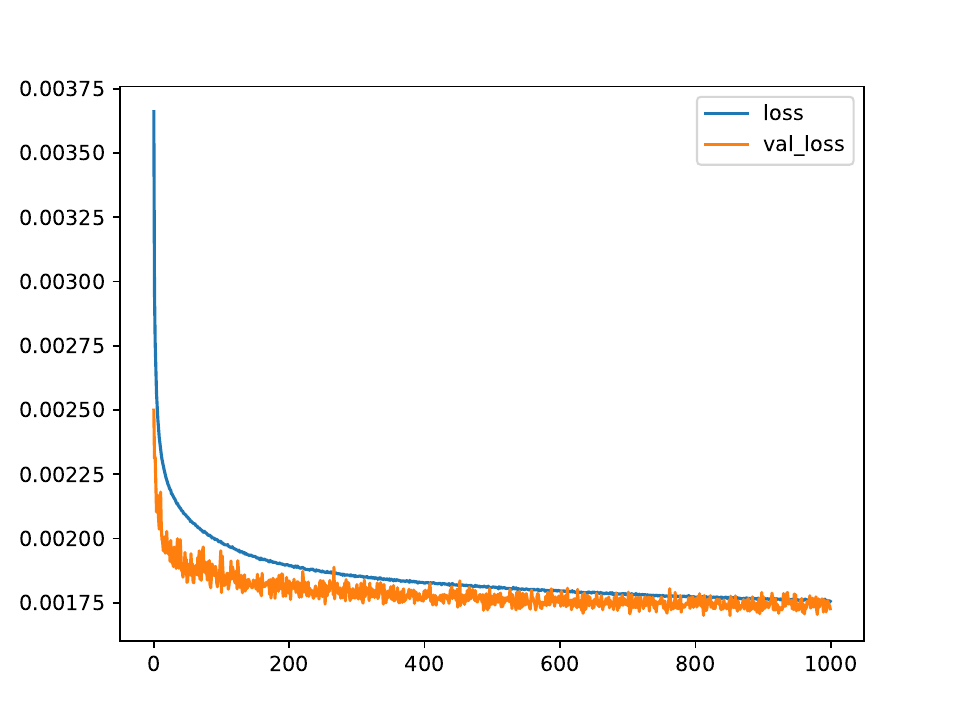}\\
(a) Guitar-only losses & (b) Augmented losses \\
\end{tabular}
\caption{Training losses for all training epochs in the guitar-only and augmented transcription problems (see Table~\ref{tab:epochs} for optimal conditions).}
\label{fig:losses_accuracy}
\end{figure*}

Figure~\ref{fig:losses_accuracy} shows the losses and the cosine-based measured accuracy in the training and validation sets for the guitar-only and augmented datasets. One thing that can be immediately noticed is the difference in both loss and accuracy during training for the two datasets. The network, with the guitar-only dataset seems to be reaching the highest learning capacity too early. Training, in both dataset cases, is stopped at the epoch with the lowest validation set loss (orange lines in (a) and (b) graphs of Figure~\ref{fig:losses_accuracy}) to avoid overfitting. Table~\ref{tab:epochs} shows that this point is reached in epoch 59 for the guitar-only dataset and 852 for the augmented dataset. It should be noted that the plotted curves in the guitar-only case are shown up to around half the epochs in comparison to the augmented dataset case, where all 1000 epochs are shown.


\begin{table}[ht]
\centering
\begin{tabular}{ r | c c }
  & guitar data & augmented data \\ \hline
 epoch & 59 & 852 \\
 train.\ loss  & $1.029\times 10^{-3}$ & $1.767\times 10^{-3}$ \\
 val.\ loss  & $1.259\times 10^{-3}$ & $1.775\times 10^{-3}$ \\
 train.\ accuracy  & 0.941 & 0.911 \\
 val.\ accuracy  & 0.931 & 0.913 \\
\end{tabular}
\caption{Epoch, loss and validation values when optimal validation accuracy was achieved (see Figure~\ref{fig:losses_accuracy}).}
\label{tab:epochs}
\end{table}

Even though cosine similarity is a useful indication about how accurate the estimations performed by the network are, the actual tablature produced after following the analytic step is not consider. To address this issue, the test set is employed (different from the training and evaluation sets) for examining the entire system, from input to a single binary output decision. The effectiveness of the entire system is cross-examined with the network trained and running on the guitar-only and the augmented data. A possible candidate for assessing the effectiveness of the system in terms of binary output is through precision, recall and f-measure. This method, however, would not be very informative towards interpreting the system decisions in the problem at hand. One reason is that precision and recall, and thus the f-measure, are tied with a strict correspondence. Since the number of ones in the system-generated binary tablature and the ground-truth binary output are constant, for any given example the false positives will always be the same as the false negatives, i.e., any 1 value that the system fails to recognise (false negative) will always be misplaced, producing a 1 value at a wrong position (false positive). Another reason is that there is no actual ground-truth in the augmented dataset, since it was constructed by adding pitches never existed in the tablature of the dataset. Therefore, in the cases where pitches were added, it would be normal to have a maximum precision value below 1.

More useful information about how accurately the system performs are extracted through measuring the percentage of cases the output exactly matched the ground-truth data, i.e.\ the binary output is exactly the same as the target binary tablature, in how many cases there was a partial match, i.e.\ how many times at least one pitch matched a ground-truth string and fret, and in how many cases there was no match, i.e.\ no pitch was assigned to the ground-truth string and fret. It should be noted that when the augmented dataset is tested, in the cases where pitch augmentation has occurred, it is most likely that no exact match can be found. If the tablature under examination has been augmented, the system is asked to compute a tablature that is indeed different from the initial ground-truth tablature, since additional pitches have been added and requested to be included in the output tablature. This means that the correct system response will be to produce a tablature that accommodates all requested pitches, given that they can be accommodated within the "\textbf{playable}" margins (as defined in step~\ref{item:playable} in Section~\ref{subsec:prob2actual}). Even in the case of examining an augmented binary fretboard, it could be actually possible to end up with a set of pitches for transcription that is the same with the set of pitches in the initial, non-augmented, fretboard, if the pitch reduction process in step~\ref{item:combinations} happens to lead to the set of pitches included in the initial fretboard; this, however, would be rare.


\begin{table*}[ht]
\centering
\begin{tabular}{ r | c c c c c c | c || c c c c c c | c }
 num. p.:  &	1 &	2 &	3 &	4 &	5 &	6 &	sum & 1 &	2 &	3 &	4 &	5 &	6 &	sum \\ \hline
 \multicolumn{8}{c ||}{Guitar-only training - Guitar-only test} & \multicolumn{7}{c}{Augmented training - Guitar-only test} \\ \hline
 no match  &	0.45 &	0.05 &	0.03 &	0.01 &	0.01 &	0.0 &	0.55 & 0.36 &	0.04 &	0.02 &	0.01 &	0.01 &	0.0 &	0.44 \\
 partial  &	0.0 &	0.03 &	0.05 &	0.01 &	0.0 &	0.0 &	0.09 & 0.0 &	0.03 &	0.03 &	0.01 &	0.0 &	0.0 &	0.07 \\
 match   &	0.25 &	0.05 &	0.03 &	0.01 &	0.0 &	0.01 &	0.35 & 0.34 &	0.06 &	0.05 &	0.01 &	0.01 &	0.01 &	0.48 \\
 sum     &	0.7 &	0.13 &	0.11 &	0.03 &	0.02 &	0.01 &	& 0.7 &	0.13 &	0.11 &	0.03 &	0.02 &	0.01 &	 \\ \hline \hline
  \multicolumn{8}{c ||}{Guitar-only training - Augmented test} & \multicolumn{7}{c}{Augmented training - Augmented test} \\ \hline
 no match  &	0.16 &	0.16 &	0.02 &	0.03 &	0.03 &	0.05 &	0.45 & 0.12 &	0.15 &	0.02 &	0.03 &	0.03 &	0.05 &	0.40 \\
 partial  &	0.0 &	0.13 &	0.06 &	0.07 &	0.06 &	0.07 &	0.39 & 0.0 &	0.15 &	0.06 &	0.07 &	0.06 &	0.08 &	0.42 \\
 match   &	0.13 &	0.01 &	0.0 &	0.0 &	0.0 &	0.0 &	0.14 & 0.17 &	0.01 &	0.0 &	0.0 &	0.0 &	0.0 &	0.18 \\
 sum     &	0.29 &	0.31 &	0.09 &	0.1 &	0.09 &	0.12 &	& 0.29 &	0.31 &	0.09 &	0.1 &	0.09 &	0.12 &	 \\
\end{tabular}
\caption{Cross-comparison results when model was trained and tested with guitar-only and augmented data.}
\label{tab:comparison}
\end{table*}

Table~\ref{tab:comparison} shows the cross-comparison between the two versions of the system, i.e.\ trained with guitar-only and augmented data. Both versions are examined by running on the both the guitar-only and the augmented test sets, in 5000 random tablature frame assignment tasks. Information in this table is given for all number of pitches on the fretboard, to show that in the guitar-only test dataset there is a clear dominance (70\%) of single-pitch instances to be transcribed. This dominance fades out with the augmentation process, since 50\% of those single-pitch instances obtain a second, added pitch. It is clear that the system trained with the augmented dataset performs better not only when tested on the augmented dataset (62\% vs 55\% partial plus absolute matches), but also on guitar-only dataset (55\% vs 44\% partial plus absolute matches). The fact that both versions perform better in the augmented rather than the guitar-only test sets, can be mainly attributed to the fact that the guitar-only dataset includes much more single-note instance, where both versions exhibit exceptionally bad performance (especially the guitar-only trained version has 45\% failures and 25\% successes).


\begin{figure*}[ht]
  \centering
\begin{tabular}{cc}
\includegraphics[width=0.49\textwidth]{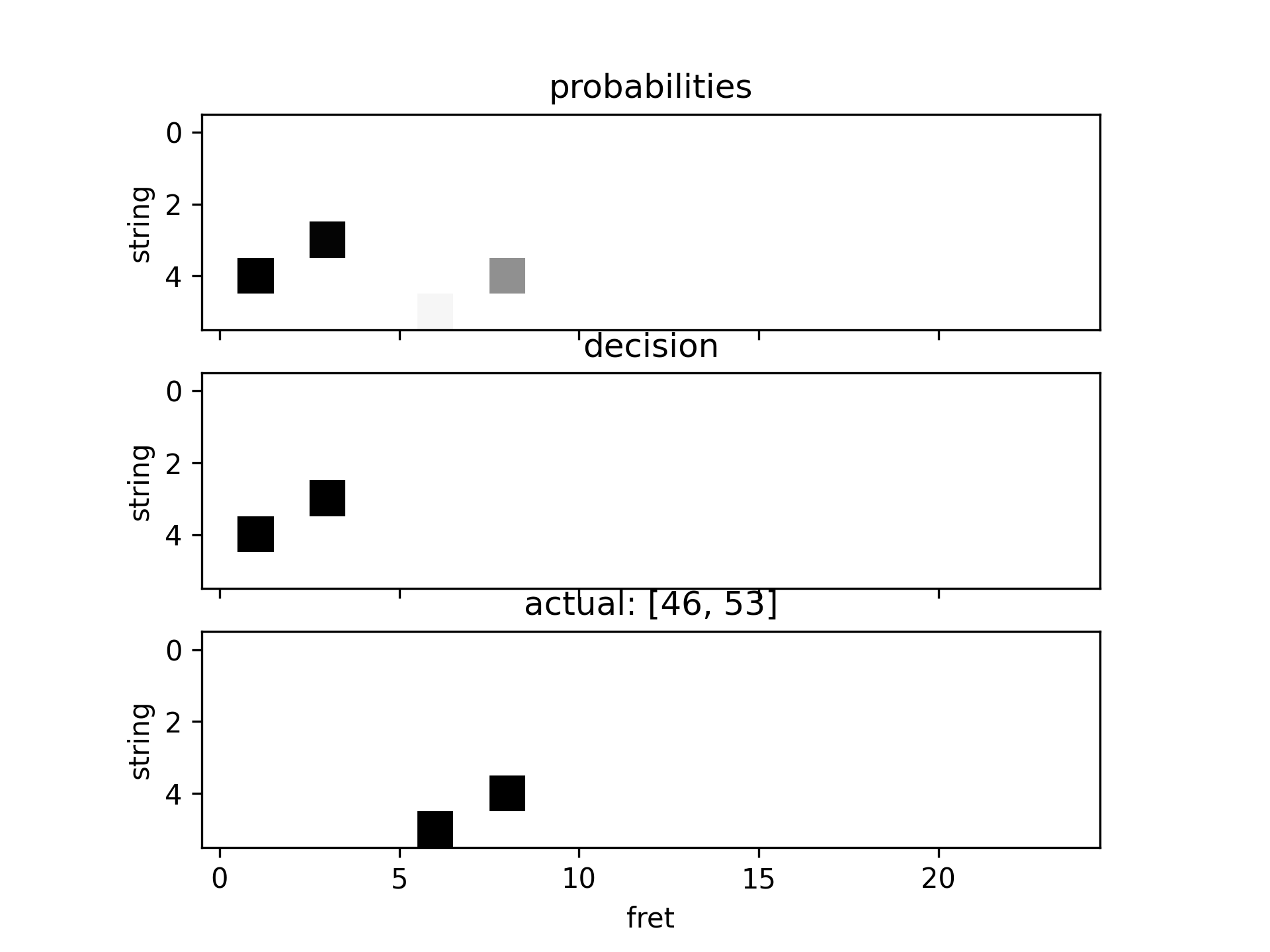} & \includegraphics[width=0.49\textwidth]{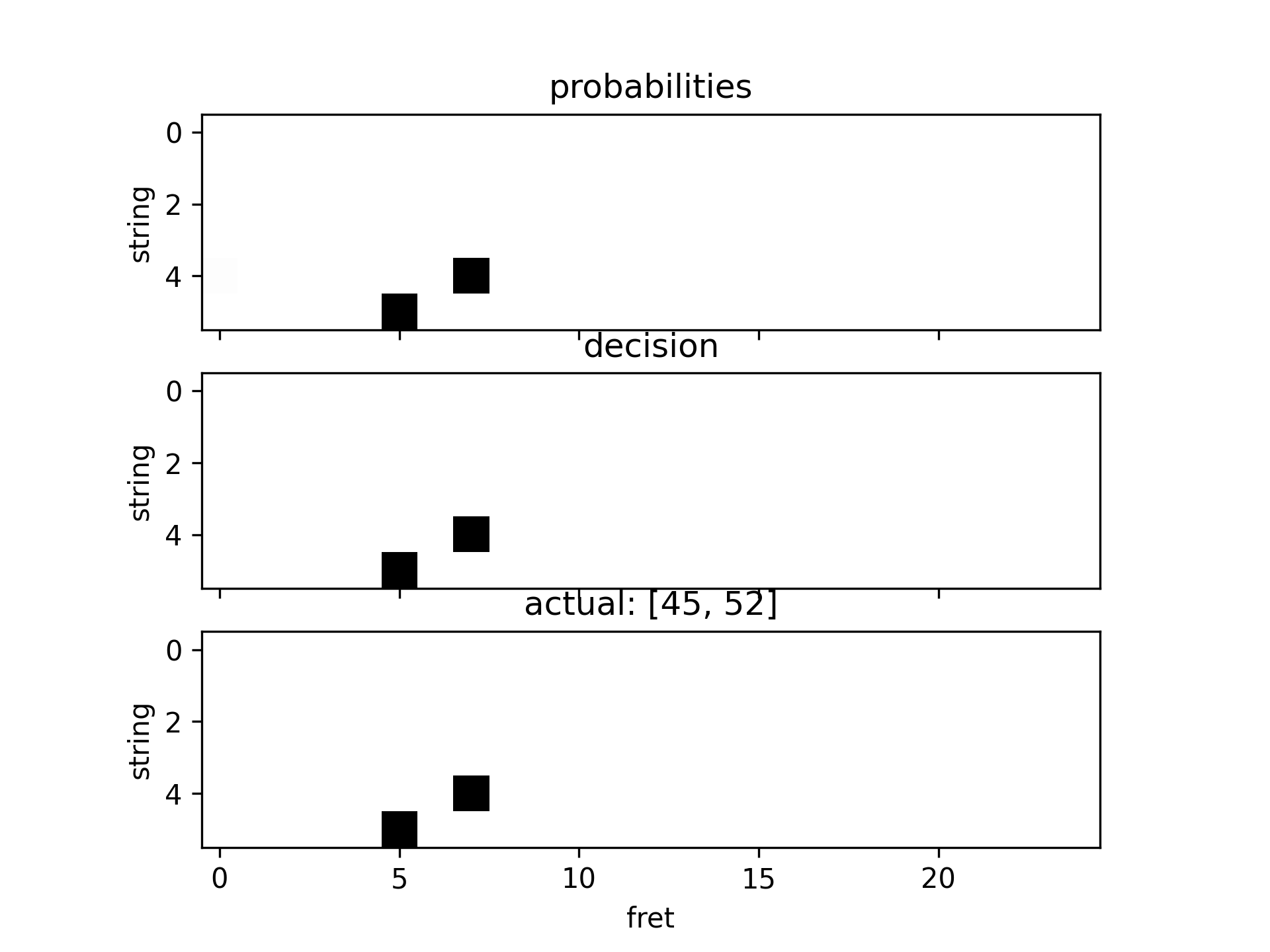}\\
(a) t-1 & (b) t \\
\end{tabular}
\caption{Example that indicates the possible usefulness of incorporating future information.}
\label{fig:bidirectionality}
\end{figure*}

A problem that has been identified with the proposed method, is that fact that sometimes is would be helpful to incorporate future information for deciding the tablature in the current step. For instance, Figure~\ref{fig:bidirectionality}~(a) shows that the decision did not match the ground-truth at time step t-1. If the system had the chance to ``look'' at the next time step, t, shown in (b), it would be more obvious to make a decision that is identical to the ground truth. Even in this version of the system, where only past information is employed, there is a weak activation in the probabilistic tablature (top graph in (a)) on one of the notes of the ground truth. A different architecture, that would incorporate future information could be able to assign stronger probabilities in string-fret positions that correspond to the actual ground truth. How this architecture would be trained or what methods could be used to make such an architecture generate a probabilistic fretboard is left for future work.

\begin{figure}[h]
\centering
\includegraphics[width=0.49\textwidth]{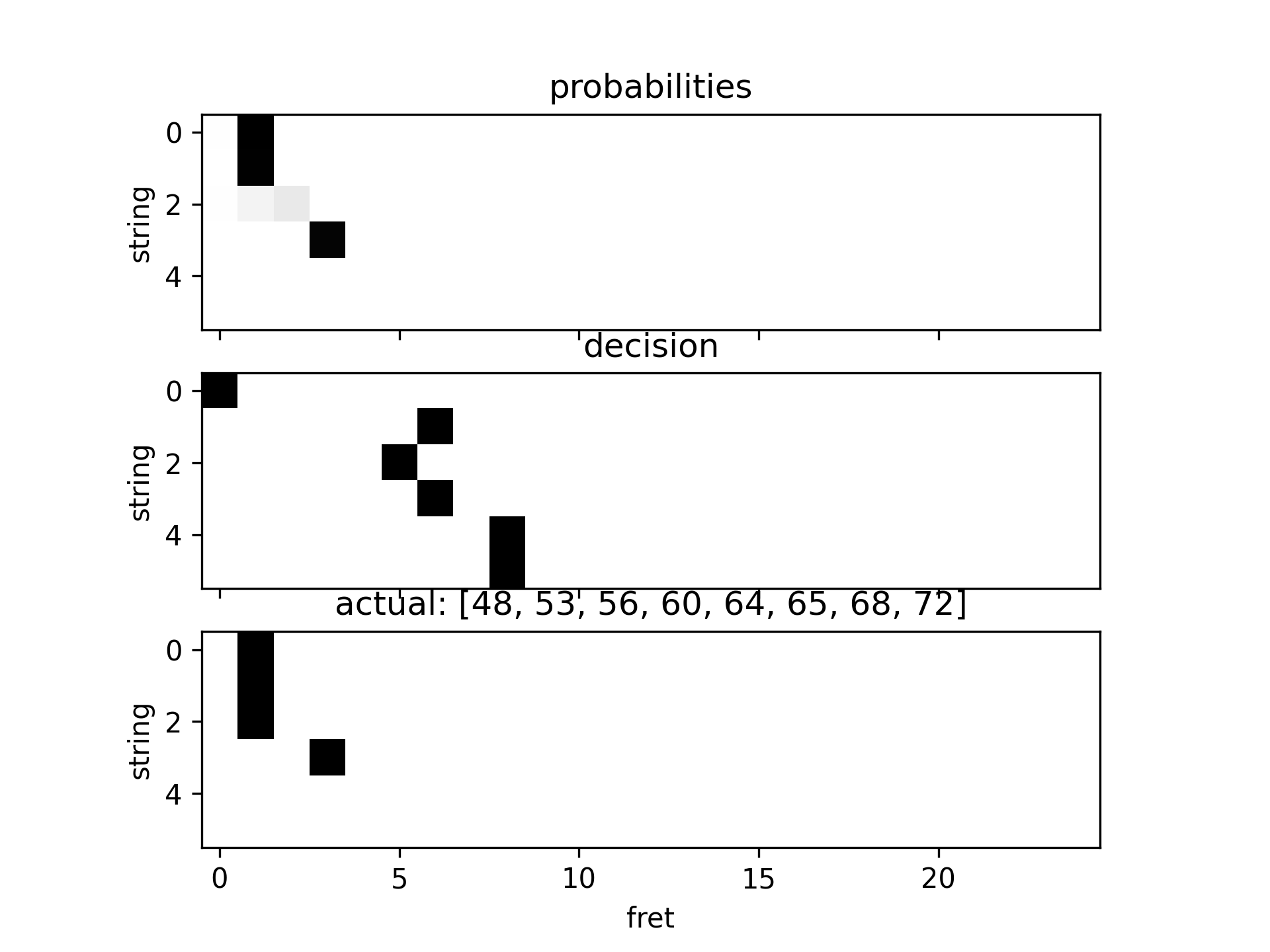}
\caption{Problem with maximal pitch coverage requirement.\label{fig:aug_problem}}
\end{figure}

Another problem that has been observed concerns the analytical step in the augmented dataset tests. This step has been designed to be ``greedy'', i.e., it starts by trying to fit as much of the requested pitches on the fretboard as possible. For example, if more than size pitches are requested, this step will first generate and examine all combinations that comprise playable solution of six pitches, regardless of how probable (as a fingering in the dataset) they are. If such a playable combination is identified, even with a near-zero probability (according to eq.~\ref{eq:analytic}), this will be returned as a solution. In some cases, this strategy is not optimal; such an example is illustrated in Figure~\ref{fig:aug_problem}. The caption of the bottom graph (``actual'' pitches) shows that eight pitches are requested and the system is able to accommodate six of them in a playable combination on the fretboard (middle graph). The probabilities that correspond to those fretboard pitches, however, are close to zero, as indicated by the top graph, which shows the network probabilistic tablature output. In fact, the network has been trained so efficiently, that regardless of the fact that eight pitches are requested in the input, the probabilistic tablature output follows a clear four-string major or minor pattern that is very close to the pre-augmentation ground-truth tablature (bottom graph). Therefore, in this case, it would possibly be better for the analytical step to continue examining fingerings with less than six pitches, even though such a fingering was found, since fingerings with less pitches could, in fact, lead to tablatures with higher probabilities.

%% file: 4_conclusions.tex
\section{Conclusions}\label{sec:conclusions}

This paper presents a machine learning method for transcribing midi pitch information to guitar tablature. The method is based on neural networks that, at a final stage, perform deconvolutions; a part of the DadaGP was employed for learning and examination purposes. Transcription of pitches that cannot possibly be played by a guitar was also examined, through a method that augments data artificially. Results indicate that training on artificial data, even though problematic in specific cases, generally led to better performance, even in single-pitch cases, where augmentation could not have happened. Based on the results, the analytic part of the method should be improved, so that fingerings that do not necessarily incorporate the maximum number of pitches can be accepted. Additionally, among the first improvements that should be attempted to the core mechanism of the method, is to incorporate information not only about the past, but also about possible futures. This could be performed either by employing bidirectional LSTM or transformer components in the architecture. 

Another improvement to the method would incorporate the examination of octave alteration of pitches that are not playable, instead of simply rejecting them. This would allow the accommodation of octave-equivalent information for as many pitches as possible in the transcription. However, such a process would require further ``stream'' information of octave-transposed pitches, a fact that would require identification of the melodic stream~\cite{cambouropoulos2008voice} that these pitches belong, along with examination of octave transposition to all the pitches within this stream. This is a complicated issue with implications in both the probabilistic and combinatorial parts of the proposed method, but the results would be much more valuable for transcribing pieces that have not been composed for and/or are not playble with guitar. Application-level improvements could include the possibility for selecting other guitar tunings, or generating exercises on given pitches with different fingering characteristics. 


%% file: main.bbl
\begin{thebibliography}{10}
\providecommand{\url}[1]{#1}
\csname url@samestyle\endcsname
\providecommand{\newblock}{\relax}
\providecommand{\bibinfo}[2]{#2}
\providecommand{\BIBentrySTDinterwordspacing}{\spaceskip=0pt\relax}
\providecommand{\BIBentryALTinterwordstretchfactor}{4}
\providecommand{\BIBentryALTinterwordspacing}{\spaceskip=\fontdimen2\font plus
\BIBentryALTinterwordstretchfactor\fontdimen3\font minus
  \fontdimen4\font\relax}
\providecommand{\BIBforeignlanguage}[2]{{%
\expandafter\ifx\csname l@#1\endcsname\relax
\typeout{** WARNING: IEEEtran.bst: No hyphenation pattern has been}%
\typeout{** loaded for the language `#1'. Using the pattern for}%
\typeout{** the default language instead.}%
\else
\language=\csname l@#1\endcsname
\fi
#2}}
\providecommand{\BIBdecl}{\relax}
\BIBdecl

\bibitem{kojs2011notating}
J.~Kojs, ``Notating action-based music,'' \emph{Leonardo Music Journal}, pp.
  65--72, 2011.

\bibitem{barbancho2012inharmonicity}
I.~Barbancho, L.~J. Tardon, S.~Sammartino, and A.~M. Barbancho,
  ``Inharmonicity-based method for the automatic generation of guitar
  tablature,'' \emph{IEEE Transactions on Audio, Speech, and Language
  Processing}, vol.~20, no.~6, pp. 1857--1868, 2012.

\bibitem{michelson2018automatic}
J.~Michelson, R.~Stern, and T.~Sullivan, ``Automatic guitar tablature
  transcription from audio using inharmonicity regression and bayesian
  classification,'' in \emph{Audio Engineering Society Convention 145}.\hskip
  1em plus 0.5em minus 0.4em\relax Audio Engineering Society, 2018.

\bibitem{hjerrild2019estimation}
J.~M. Hjerrild and M.~G. Christensen, ``Estimation of guitar string, fret and
  plucking position using parametric pitch estimation,'' in \emph{ICASSP
  2019-2019 IEEE International Conference on Acoustics, Speech and Signal
  Processing (ICASSP)}.\hskip 1em plus 0.5em minus 0.4em\relax IEEE, 2019, pp.
  151--155.

\bibitem{abesser2012automatic}
J.~Abe{\ss}er, ``Automatic string detection for bass guitar and electric
  guitar,'' in \emph{International Symposium on Computer Music Modeling and
  Retrieval}.\hskip 1em plus 0.5em minus 0.4em\relax Springer, 2012, pp.
  333--352.

\bibitem{kehling2014automatic}
C.~Kehling, J.~Abe{\ss}er, C.~Dittmar, and G.~Schuller, ``Automatic tablature
  transcription of electric guitar recordings by estimation of score-and
  instrument-related parameters.'' in \emph{DAFx}, 2014, pp. 219--226.

\bibitem{barbancho2011automatic}
A.~M. Barbancho, A.~Klapuri, L.~J. Tard{\'o}n, and I.~Barbancho, ``Automatic
  transcription of guitar chords and fingering from audio,'' \emph{IEEE
  Transactions on Audio, Speech, and Language Processing}, vol.~20, no.~3, pp.
  915--921, 2011.

\bibitem{burlet2013robotaba}
G.~Burlet and I.~Fujinaga, ``Robotaba guitar tablature transcription
  framework.'' in \emph{ISMIR}, 2013, pp. 517--522.

\bibitem{yazawa2013audio}
K.~Yazawa, D.~Sakaue, K.~Nagira, K.~Itoyama, and H.~G. Okuno, ``Audio-based
  guitar tablature transcription using multipitch analysis and playability
  constraints,'' in \emph{2013 IEEE International Conference on Acoustics,
  Speech and Signal Processing}.\hskip 1em plus 0.5em minus 0.4em\relax IEEE,
  2013, pp. 196--200.

\bibitem{yazawa2014automatic}
K.~Yazawa, K.~Itoyama, and H.~G. Okuno, ``Automatic transcription of guitar
  tablature from audio signals in accordance with player's proficiency,'' in
  \emph{2014 IEEE International Conference on Acoustics, Speech and Signal
  Processing (ICASSP)}.\hskip 1em plus 0.5em minus 0.4em\relax IEEE, 2014, pp.
  3122--3126.

\bibitem{sayegh1989fingering}
S.~I. Sayegh, ``Fingering for string instruments with the optimum path
  paradigm,'' \emph{Computer Music Journal}, vol.~13, no.~3, pp. 76--84, 1989.

\bibitem{radisavljevic2004path}
A.~Radisavljevic and P.~F. Driessen, ``Path difference learning for guitar
  fingering problem,'' in \emph{ICMC}, vol.~28, 2004.

\bibitem{radicioni2005guitar}
D.~Radicioni and V.~Lombardo, ``Guitar fingering for music performance,'' in
  \emph{ICMC}, 2005.

\bibitem{miura2004constructing}
M.~Miura, I.~Hirota, N.~Hama, and M.~Yanagida, ``Constructing a system for
  finger-position determination and tablature generation for playing melodies
  on guitars,'' \emph{Systems and Computers in Japan}, vol.~35, no.~6, pp.
  10--19, 2004.

\bibitem{hori2013input}
G.~Hori, H.~Kameoka, and S.~Sagayama, ``Input-output hmm applied to automatic
  arrangement for guitars,'' \emph{Information and Media Technologies}, vol.~8,
  no.~2, pp. 477--484, 2013.

\bibitem{tuohy2005genetic}
D.~R. Tuohy and W.~D. Potter, ``A genetic algorithm for the automatic
  generation of playable guitar tablature,'' in \emph{ICMC}, 2005, pp.
  499--502.

\bibitem{tuohy2006ga}
------, ``Ga-based music arranging for guitar,'' in \emph{2006 IEEE
  International Conference on Evolutionary Computation}.\hskip 1em plus 0.5em
  minus 0.4em\relax IEEE, 2006, pp. 1065--1070.

\bibitem{tuohy2006evolved}
D.~R. Tuohy, W.~D. Potter, and A.~I. Center, ``An evolved neural network/hc
  hybrid for tablature creation in ga-based guitar arranging,'' in
  \emph{ICMC}.\hskip 1em plus 0.5em minus 0.4em\relax Citeseer, 2006.

\bibitem{tuohy2006generating}
D.~R. Tuohy and W.~D. Potter, ``Generating guitar tablature with lhf notation
  via dga and ann,'' in \emph{International Conference on Industrial,
  Engineering and Other Applications of Applied Intelligent Systems}.\hskip 1em
  plus 0.5em minus 0.4em\relax Springer, 2006, pp. 244--253.

\bibitem{tuohy2006guitar}
D.~R. Tuohy and W.~Potter, ``Guitar tablature creation with neural networks and
  distributed genetic search,'' in \emph{Proc. of the 19th International
  Conference on Industrial and Engineering Applications of Artificial
  Intelligence and Expert Systems, IEA-AIE06, Annecy, France}, 2006.

\bibitem{ramos2015comparative}
J.~V. Ramos, A.~S. Ramos, C.~N. Silla, and D.~S. Sanches, ``Comparative study
  of genetic algorithm and ant colony optimization algorithm performances for
  the task of guitar tablature transcription,'' in \emph{2015 Brazilian
  Conference on Intelligent Systems (BRACIS)}.\hskip 1em plus 0.5em minus
  0.4em\relax IEEE, 2015, pp. 228--233.

\bibitem{ramos2016evaluation}
------, ``An evaluation of different evolutionary approaches applied in the
  process of automatic transcription of music scores into tablatures,'' in
  \emph{2016 IEEE 28th International Conference on Tools with Artificial
  Intelligence (ICTAI)}.\hskip 1em plus 0.5em minus 0.4em\relax IEEE, 2016, pp.
  663--669.

\bibitem{gagnon2003neural}
T.~Gagnon, S.~Larouche, and R.~Lefebvre, ``A neural network approach for
  preclassification in musical chords recognition,'' in \emph{The
  Thrity-Seventh Asilomar Conference on Signals, Systems \& Computers, 2003},
  vol.~2.\hskip 1em plus 0.5em minus 0.4em\relax IEEE, 2003, pp. 2106--2109.

\bibitem{humphrey2014music}
E.~J. Humphrey and J.~P. Bello, ``From music audio to chord tablature: Teaching
  deep convolutional networks toplay guitar,'' in \emph{2014 IEEE international
  conference on acoustics, speech and signal processing (ICASSP)}.\hskip 1em
  plus 0.5em minus 0.4em\relax IEEE, 2014, pp. 6974--6978.

\bibitem{wiggins2019guitar}
A.~Wiggins and Y.~Kim, ``Guitar tablature estimation with a convolutional
  neural network.'' in \emph{ISMIR}, 2019, pp. 284--291.

\bibitem{chenautomatic}
Y.-H. Chen, Y.-H. Huang, W.-Y. Hsiao, and Y.-H. Yang, ``Automatic composition
  of guitar tabs by transformers and groove modeling.''

\bibitem{sarmento2021dadagp}
P.~Sarmento, A.~Kumar, C.~Carr, Z.~Zukowski, M.~Barthet, and Y.-H. Yang,
  ``Dadagp: A dataset of tokenized guitarpro songs for sequence models,''
  \emph{arXiv preprint arXiv:2107.14653}, 2021.

\bibitem{cambouropoulos2008voice}
E.~Cambouropoulos, ``Voice and stream: Perceptual and computational modeling of
  voice separation,'' \emph{Music Perception}, vol.~26, no.~1, pp. 75--94,
  2008.

\end{thebibliography}
